\documentclass[sigconf]{acmart}

\usepackage{listings, xcolor, enumitem, caption}

\AtBeginDocument{%
  }

\setcopyright{acmcopyright}
\copyrightyear{2018}
\acmYear{2018}
\acmDOI{XXXXXXX.XXXXXXX}

\acmConference[Conference acronym 'XX]{Make sure to enter the correct
  conference title from your rights confirmation emai}{June 03--05,
  2018}{Woodstock, NY}
\acmPrice{15.00}
\acmISBN{978-1-4503-XXXX-X/18/06}

\newcommand{\mynote}[3]{}

\newcommand{\ra}[1]{\mynote{Ravil}{#1}{red}}

\begin{document}

\title{Benchmark Framework with Skewed Workloads}

\author{Ravil Galiev}
\email{ravil.galiev.2000@gmail.com}
\affiliation{%
  \institution{ITMO University}
  \country{Russia}
}

\author{Dmitry Ivanov}
\email{dmitry.ivanov@huawei.com}
\affiliation{%
  \institution{Huawei}
  \country{Russia}
}

\author{Vitaly Aksenov}
\email{aksenov.vitaly@gmail.com}
\affiliation{%
  \institution{ITMO University}
  \country{Russia}
}

\renewcommand{\shortauthors}{Galiev R. et al.}

\begin{abstract}
In this work, we present a new benchmarking suite with new real-life inspired skewed workloads to test the performance of concurrent index data structures. We started this project to prepare workloads specifically for self-adjusting data structures, i.e., they handle more frequent requests faster, and, thus, should perform better than their standard counterparts. We looked over the commonly used suites to test performance of concurrent indices trying to find an inspiration: Synchrobench, Setbench, YCSB, and TPC~--- and we found several issues with them.

The major problem is that they are not flexible: it is difficult to introduce new workloads, it is difficult to set the duration of the experiments, and it is difficult to change the parameters. We decided to solve this issue by presenting a new suite based on Synchrobench.

Finally, we highlight the problem of measuring performance of data structures. We show that the relative performance of data structures highly depends on the workload: it is not clear which data structure is best. For that, we take three state-of-the-art concurrent binary search trees and run them on the workloads from our benchmarking suite. As a result, we get six experiments with all possible relative performance of the chosen data structures.
\end{abstract}



\maketitle

\section{Introduction}
One of the largest open questions is how to understand which particular data structure works better than others. The standard approach in research papers (or in practical research) is to run the data structures on a specific set of workloads. If the target data structure outperforms others, i.e., executes all the operations faster on all the chosen workloads, then one can claim that this structure is indeed better.

In this work, we consider concurrent index data structures that store keys with values. They support three operations: 1)~\texttt{insert} operation adds a key with the provided value to the set and returns a previous corresponding value; 2)~\texttt{remove} operation removes a key from the set and returns the corresponding value; 3)~\texttt{get} operation returns a value that corresponds to a requested key. The operations \texttt{insert} and \texttt{remove} can be referred to as write or update operations, and \texttt{get} operation can be referred to as a read operation.

There exist several benchmarking suites to test concurrent index data structures. The most popular of them are: Synchrobench~\cite{gramoli2015more}, Setbench~\cite{brown2020non}, YCSB~\cite{cooper2010benchmarking}, and TPC benchmarks~\cite{TPC}. Let us take a deeper look on them. Synchrobench is the simplest one: it provides only an uniform workload. In the uniform workload, threads in an infinite loop chooses a type of an operation corresponding to the probability of write and read operations and as an argument it takes a key chosen uniformly at random from the specified range. Synchrobench is implemented in Java and C++ and provides a lot of state-of-the-art implementations.

Setbench is more generic. In addition to the Uniform workload, it provides the Zipfian one~--- where the arguments of the operations are taken from Zipfian distribution~\cite{powers1998applications}. Moreover, it provides one workload from both YCSB and TPC-C. Setbench is written in C++ and provides a lot of state-of-the-art implementations.

Now, we come to the industrially used benchmarking suites: YCSB and TPC.
YCSB provides five predefined workloads~\cite{cooper2010benchmarking}:
\begin{itemize}
\item update heavy (A): $50\%$ of updates and $50\%$ of reads, Zipfian distribution of arguments;
\item read heavy (B): $5\%$ of updates and $95\%$ of reads, Zipfian distribution of arguments;
\item read only (C): $100\%$ of reads, Zipfian distribution of arguments;
\item read latest (D): $5\%$ of inserts and $95\%$ of reads, Latest with Zipfian distribution of arguments, i.e., the recently accessed \ra{inserted} keys have bigger probability;
\item and, scan workload (E): $95\%$ of operations are scans, i.e., traverse through all elements from a specified range, and $5\%$ of inserts with Zipfian and Uniform distributions of arguments.
\end{itemize}

TPC is a meta-suite, i.e., a suite of suites. Its workloads are more complicated than YCSB workloads since they use transactions in the database which we currently do not support. However, they are more practically in nature. The subsuites of our interest are TPC-C and TPC-H. We omit their explanation here. However, they do not satisfy our goals and we later explain~--- why.

Now, we state four reasonable requirements for the benchmarking suites that we expect from them. At first, it should be easy to add a new workload to the suite. Secondly, we want workloads to run for an arbitrary time, i.e., we require infinite workloads. Thirdly, or the most specific, we want the workloads to be: 1)~taken from the real-life patterns; 2)~be skewed, so that we can use them to test self-adjusting data structures, i.e., data structures that handle more frequent requests faster. Finally, we want to tune workload parameters in a simple manner.

We consider each requirement one by one. For the first one, none of the benchmarking suites satisfy us~--- you have to write your workload totally from scratch including all the necessary functions, such as the logic with the threads and their execution, the reading of the arguments, the calculation of the statistics, etc. We find it very tedious~--- we tried to fix it by providing a new modular approach to a suite that should require only to rewrite small parts of the benchmark. In the main part, we present how easy it is to add the skewed workload presented in~\cite{aksenov2023splay} to our suite.

For the second requirement, the Uniform and Zipfian workloads from Synchrobench and Setbench satisfy us~--- they can be repeated for an arbitrary amount of time. However, YCSB and TPC workloads have a predefined number of operations before the run. For example, consider the Latest distribution in YCSB: a thread requests a key from the latest requested ones. Unfortunately, we cannot choose an argument from this distribution in constant time during the run~--- you need at least $\Omega(\log n)$ additional operations that could affect the whole benchmark. Thus, the infinite workload seems to be impossible to implement. The same happens with TPC benchmarks. In our benchmarking suite, workloads are infinite. As an example, we present an infinite workload similar to the Latest distribution in YCSB. See Section~\ref{Creakers_and_Wave_KeyGenerator}.

Thirdly, Synchrobench and Setbench workloads are synthetic and, thus, are suitable only for microbenchmarking. YCSB is partly synthetic as our suite~--- the workloads try to model the real-life behaviour. With TPC everything gets more complicated: it uses some real-life traces with synthetic ones. For example, TPC-C uses two distributions: Uniform and NURandom (non-uniform random). We find this to be a little bit restrictive, thus, we decided to come up with new interesting workloads related to real-life. Currently, we implemented two: 1)~Temporary Skewed Sets~--- the workload depends on the current time, i.e., one wants to know the weather in the morning, to search for a work-related data in the afternoon, and to read the news in the evening; and 2)~Creakers and Wave~--- the recently added elements are requested more frequently than the old ones, i.e., musical hits are listened to more often when they have just been released. We explain these workloads in detail in Section~\ref{sec:generators}.

Finally, all the workloads in YCSB and TPC seem to be fixed: it is really difficult to change the parameters~--- there are many complicated dependencies between them. Of course, it is not the problem with the simplest Uniform and Zipfian workloads in Synchrobench and Setbench. We believe that the implemented workloads in our benchmarking suite are simpler to setup.

At the end, we state a \textbf{very important observation} concerning the benchmarking processes. The presented results in papers, in general, are not necessarily fair: typically, the authors use only the ``good'' workloads where their data structure outperforms others. We remind everyone to take it seriously--- it is hard to win on all possibly workloads. Thus, all the performance results should be considered with suspicion.

To show that, we took three well-known Java implementations of binary search trees: BCCO tree~\cite{bronson2010practical}, Concurrency-Friendly tree~\cite{crain2013contention}, and Concurrency-Optimal tree~\cite{aksenov2017concurrency}; and presented six workloads from our suite that show any possible relative performance between these three data structures.

We want to end the introduction with a small discussion of our benchmark suite. The suite currently offers several synthetic workloads out of the box, some of which are based on real-life patterns. Right now, we have only the Java version of the suite built on Synchrobench, but the C++ version based on Setbench is coming soon. One can check out the Java suite by the link~\url{https://github.com/Mr-Ravil/synchrobench}.

\textbf{Roadmap. } In Section~\ref{SOFTWARE_DESIGN}, we present a software design of our suite and provide an example on how one can add a new workload. In Sections~\ref{sec:distributions},~\ref{sec:generators}, and~\ref{sec:threadloops}, we describe the predefined entities in our suite which one can already use to build more complicated workloads. In Section~\ref{sec:separator}, we present that we really cannot separate three well-known binary search trees. Finally, we conclude in Section~\ref{sec:conclusion}.

\vspace{-0.3cm}
\section{Software Design} \label{SOFTWARE_DESIGN}

There are four types of entities in our benchmarking suite:
\vspace{-0.2cm}
\begin{itemize}
  \item \texttt{Distribution} represents a distribution of a random variable;
  \item \texttt{KeyGeneratorData} converts a generated value from a distribution, into a key;
  \item \texttt{KeyGenerator} generates a key using several Distribution and KeyGeneratorData entities;
  \item \texttt{ThreadLoop} express the logic on how to interact with a provided index data structure.
\end{itemize}

\vspace{-0.2cm}

Typically, each benchmark depends on one ThreadLoop, that uses one KeyGenerator, whereas the KeyGenerator utilizes several Distributions and KeyGeneratorDatas.

Each workload in our suite have several obligatory parameters:

\vspace{-0.2cm}
\begin{itemize}
    \item a working range of keys;
    \item a size of a data structure before the benchmark;
    \item a number of working threads during the benchmark;
    \item a number of threads at the prefill phase;
    \item a duration of the benchmarking process.
\end{itemize}

\vspace{-0.3cm}

\subsection{Distributions}

We start with the low-level entity~--- Distribution. A distribution is used to simulate some random variable. It is important to note that it generates some value from a distribution that later translates into a key. The KeyGenerator later uses this value in its calculations of the next key. Distribution provides a method \texttt{next()} that generates a value of a random variable, typically restricted to some range. There exists an extended version of Distribution~--- MutableDistribution. Its main difference is that MutableDistribution can change the random variable in progress by modifying the range of keys: \texttt{setRange(int range)} and \texttt{next(int range)}.

To create a Distribution object one could use a DistributionsBuilder. Please, see an example below.

\vspace{-0.3cm}

\subsection{KeyGeneratorData}

KeyGeneratorData does not have a single standard. Typically, it maintains an array of shuffled keys from the range. Then, KeyGenerator gets a key by feeding a value (or an index) generated by Distribution to our KeyGeneratorData.

\subsection{KeyGenerators}

KeyGenerator should be able to generate keys for each type of operation. It implements four methods: \texttt{nextGet()}, \texttt{nextInsert()}, \texttt{nextRemove()}, and \texttt{nextPrefill()}. The last operation is used to prefill the index data structure before applying operations in a ThreadLoop \texttt{prefill()} method. This prefill operation is important for KeyGenerators with specific needs.

In order to simplify the creation of a new KeyGenerator object, one can use a KeyGeneratorBuilder.
It generates a KeyGenerator for each thread running a ThreadLoop.

\vspace{-0.2cm}
\subsection{ThreadLoops}

ThreadLoop is an infinite loop that decides which operation to execute next. It is created for each thread separately and, thus, uses the corresponding KeyGenerator. ThreadLoop has two main methods: \texttt{prefill()} and \texttt{run()}. The first method explains how to prefill a data structure before benchmarking. The second method explains how to choose operations and perform it during the main phase. Default ThreadLoop randomly chooses the next operation using the uniform distribution, receives the key for the selected operation from the KeyGenerator, and executes it. Of course, one can implement their own version of ThreadLoop by extending the ThreadLoopAbstract class.

Note that ThreadLoop also calculates different statistics, for example, the number of successful operations. This is important for the benchmark in order to compare different data structures and check their correctness.

\vspace{-0.2cm}
\subsection{Example} \label{Example}

As an example, we explain on how to implement the skewed read-update workload from~\cite{aksenov2023splay} in our suite.
We start with a brief description of that workload.

This workload is specified by five parameters $n - w - x - y - s$: 
\begin{itemize}
    \item $n$, the size of the workset of keys;
    \item $w\%$, the amount of updates performed;
    \item $x\%$ of get operations choose a key uniformly at random from a random subset of keys of proportion $y\%$, while other get operations choose a random key from the rest of the set;
    \item \texttt{insert} and \texttt{remove} operation choose a key uniformly at random from a random subset of keys of proportion $s\%$.
\end{itemize}

We explain what these parameters mean in the terms of the entities described earlier.
The first parameter, $n$, mentioned in section~\ref{SOFTWARE_DESIGN}, is the range of keys. 
The second parameter, $w$, is the probability to choose an \texttt{update} (\texttt{insert} and \texttt{erase}) operation in the ThreadLoop.
Parameters $x$ and $y$ affect three entities at once: Distribution, KeyGeneratorData, and KeyGenerator.
For \texttt{get} operation, the Distribution generates a random variable from $[0; y \cdot n)$ uniformly with a probability of $x\%$ and a random variable from $[y \cdot n, n)$ uniformly with a probability of $100 - x\%$. We name this distribution the Skewed Uniform Distribution. As a KeyGeneratorData we take just a shuffled array. Using these Distribution and Data, the KeyGenerator generates a key for \texttt{get} operations.
Finally, in a similar manner, we can explain a parameter $s$.


Now, we give a more detailed description of all entities used in this workload.


\subsubsection{Distribution} \label{SKEWED_UNIFORM}
In this workload we need two distributions: a Uniform Distribution and a Skewed Uniform Distribution explained above. The Uniform Distribution is quite simple and is available out of the box in the benchmarking suite. Thus, we explain on how to implement the Skewed Uniform Distribution. Together with Data, it is used to generate the keys for get operations in the KeyGenerator.

To start with, we need to add the distribution type to the DistributionType enum:
\begin{lstlisting}[language = C++]
enum DistributionType:
  {...}, SKEWED_UNIFORM
\end{lstlisting}


Then, we need to provide parameters $x\%$ and $y\%$ to this distribution. For that, we need to create a parameters object:
\begin{lstlisting}[language = C++]
class SkewedUniformParameters:
  double HOT_SIZE = 0 // y%
  double HOT_PROB = 0 // x%
\end{lstlisting}



The user needs to write a parser for the command line in DistributionBuilder in order to extract the parameters for our Distribution:
    

We are ready to implement SkewedUniformDistribution. Since we need to choose a value uniformly from both sets $[0, y \cdot n)$ and $[y \cdot n, n)$, corresponding to $x\%$ and $100-x\%$, one can use just two different Uniform Distribution objects.

\begin{lstlisting}[language = C++]
class SkewedUniformDistribution 
            implements Distribution:
  int hotLength
  double hotProb
  UniformDistribution hotDist
  UniformDistribution coldDist
  Random random

  int next():
    if (random.nextDouble() < hotProb):
      return hotDist.next()
    else:
      return hotLength + coldDist.next()
\end{lstlisting}





Distribution returns only a random variable value, not an operation argument. We need to use KeyGeneratorData to translate this value into a key.

To finish with the implementation of the new Distribution, the user should write a specific builder in DistributionBuilder (by default, DistributionBuilder builds the UniformDistribution).
\begin{lstlisting}[language = C++]
Distribution getDistribution(int range):
  switch (this.distributionType) {
    case SKEWED_UNIFORM:
      int hotLength = parameters.HOT_SIZE * range; 
      return new SkewedUniformDistribution(
        hotLength,
        parameters.HOT_PROB,
        new DistributionBuilder()
          .getDistribution(hotLength),
        new DistributionBuilder()
          .getDistribution(range - hotLength))
    case {...}
\end{lstlisting}


\subsubsection{KeyGenerator} \label{EXAMPLE_KEYGEN}
After the Distribution, we can implement a new KeyGenerator, EXAMPLE\_KEYGEN. We should notify the system about its existence by appending into KeyGeneratorType enum:
\begin{lstlisting}[language = C++]
enum KeyGeneratorType:
  {...}, EXAMPLE_KEYGEN
\end{lstlisting}

KeyGenerator takes three parameters: $x$, $y$, and $s$, discussed above. $x$, $y$ are used for the Skewed Uniform Distribution and $s$ is used for the Uniform Distribution.

At first, we need to provide the parameters to a KeyGenerator. We use \texttt{build()} function to set up all Distributions.
\begin{lstlisting}[language = C++]
class ExampleParameters extends Parameters:
  DistributionBuilder getDistBuilder = 
    new DistributionBuilder(SKEWED_UNIFORM)
  DistributionBuilder updateDistBuilder = 
    new DistributionBuilder(UNIFORM)
  double x, y, s

  void build():
    super.build()
    getDistBuilder.setParameters(
      new SkewedUniformParameters(x, y))
\end{lstlisting}

Then, one should implement the interface of the KeyGenerator:
\begin{lstlisting}[language = C++]
interface KeyGenerator:
    int nextRead()
    int nextInsert()
    int nextErase()
    int nextPrefill()  
\end{lstlisting}

Since the Distribution returns random variables, the KeyGenerator needs a tool to convert them into keys and that is KeyGeneratorData.
In this case, the KeyGeneratorData takes all keys from the entire range and shuffles them. Then, KeyGenerator takes the index generated by the Distribution and passes it to the KeyGeneratorData that returns the corresponding key.
In our case, we need two KeyGeneratorData for read and update operations separately.
\begin{lstlisting}[language = C++]
class ExampleKeyGenerator implements KeyGenerator: 
  KeyGeneratorData getData
  KeyGeneratorData updateData
  Distribution getDist
  Distribution updateDist

  int nextGet():
    return getData.get(getDist.next())

  int nextInsert(): 
    return updateData.get(updateDist.next())

  int nextRemove():
    return updateData.get(updateDist.next())
\end{lstlisting}

KeyGeneratorBuilder is the next step in the implementation of the KeyGenerator. The generateKeyGenerators() method creates an array of ExampleKeyGenerator, one for each thread, and initializes them:
\begin{lstlisting}[language = C++]
class ExampleKeyGeneratorBuilder 
            extends KeyGeneratorBuilder:
  KeyGenerator[] generateKeyGenerators():
    KeyGenerator[] keygens = 
      new KeyGenerator[parameters.numThreads]
    KeyGeneratorData getData = 
      new KeyGeneratorData(parameters)
    KeyGeneratorData updateData = 
      new KeyGeneratorData(parameters)

    for (i = 0; i < parameters.numThreads; i++):
      keygens[i] = new ExampleKeyGenerator(
        getData, updateData,
        parameters.getDistBuilder
          .getDistribution(parameters.range),
        parameters.upadeteDistBuilder
          .getDistribution(parameters.s *
                           parameters.range))
        
    return keygens
\end{lstlisting}




At the end, the user should add his KeyGenerator to \\ the \texttt{parseKeyGenerator()} method in Parameters class.



\subsubsection{ThreadLoop} \label{DefaultThreadLoop}
The implementation of a new ThreadLoop consists of four parts: 1)~an implementation of the ThreadLoopAbstract class, i.e., \texttt{run()} and \texttt{prefill()} operations; 2)~an implementation of the ThreadLoopParameters interface with a list of parameters and \texttt{parseArg()} operation; 3)~add the new ThreadLoop to the \texttt{parseThreadLoop()} method in Parameters class; and 4)~add the new ThreadLoop through \texttt{initThreads()} method in the main Test class.

The ThreadLoop interface exports two main methods: \texttt{run()} and \texttt{prefill()}. The \texttt{run} method is pretty simple: while the stop flag is false, the method selects a next operation, executes it, and collects statistics. The \texttt{prefill()} method takes an AtomicInteger \texttt{prefillSize} as an argument. Its value indicates how many keys are left to add. This variable is an AtomicInteger since we should allow to prefill the structure with several threads. Here is an example of its implementation:
\begin{lstlisting}[language = C++]
void prefill(AtomicInteger prefillSize):
  while (prefillSize.get() > 0):
    int curSize = prefillSize.decrementAndGet()
    int v = keygen.nextPrefill()
    if (curSize < 0 
      || bench.putIfAbsent(v, v) != null):
      prefillSize.incrementAndGet()
\end{lstlisting}

One can say that \texttt{prefill()} should be performed sequentially, not concurrently by different threads~--- here, we point to the discussion by Brown et al.~\cite {kharal2022performance}. If one still wants the prefill to be sequential, there is a possibility~--- the user need to set the number of prefilling threads to one.


\section{Implemented Distributions}
\label{sec:distributions}

In this Section, we overview Distributions provided in our benchmark out of the box.

\vspace{-0.2cm}
\subsection{Uniform Distribution}
In the Uniform Distribution, a random value is generated with equal probability from a specified range. The Uniform Distribution implements MutableDistribution interface since it does not restrict the range of values. In other words, we can generate a random value using \texttt{random.nextInt(range)}.

\vspace{-0.2cm}
\subsection{Skewed Uniform Distribution}
The Skewed Uniform Distribution was explained above, in section~\ref{SKEWED_UNIFORM}.

\vspace{-0.2cm}
\subsection{Zipfian Distribution}
The Zipfian Distribution is based on Zipf's law~\cite{powers1998applications}.

\vspace{-0.2cm}
\section{Implemented KeyGenerators}
\label{sec:generators}



In this section, all KeyGeneratorData objects generate an array of keys from the entire
range and shuffles it (\ref{EXAMPLE_KEYGEN}).
When choosing an argument a KeyGenerator sends an index, i.e., a generated value by a Distribution, to the KeyGeneratorData and gets the corresponding key. User can enable non-shuffle mode for such KeyGeneratorData implementations.

\vspace{-0.2cm}
\subsection{Default KeyGenerator}
The Default KeyGenerator accepts one of the existing distributions as input and selects the next key based on this distribution from KeyGeneratorData.

\vspace{-0.2cm}
\subsection{Skewed Sets KeyGenerator}
Now, we present the Skewed Sets KeyGenerator which is an improvement of the KeyGenerator described in section~\ref{EXAMPLE_KEYGEN}.

This KeyGenerator uses two SkewedUniformDistributions separately for read and update operations. It takes the following parameters:
\begin{itemize}
    \item $rp\%$ of read operations are performed on a random subset of keys of proportion $rs\%$ where a key is taken uniformly. All other read operations are performed on the rest of the set.
    \item $wp\%$ of update operations are performed on a random subset of keys of proportion $ws\%$ where a key is taken uniformly. All other update operations are performed on the rest of the set.
    \item $inter\%$ of keys are in the intersection of the working sets of read and update operations. 

\end{itemize}


\vspace{-0.2cm}
\subsection{Temporary Skewed Sets KeyGenerator}
The TemporarySkewedKeyGenerator is motivated by the fact that a user may want to ask frequently different subsets of keys in different periods of time. For example, one wants to know the weather in the morning, search for a work-related data in the afternoon, and read the news in the evening. To simulate this, the Temporary Skewed Sets KeyGenerator has two types of states:
\begin{itemize}
    \item $k$-th \emph{excited} state~--- the keys are selected using $k$-th SkewedUniformDistribution, i.e., there is a hot set of keys.
    \item a \emph{dormant} state~--- all keys are selected with the UniformDistribution. This state happens between $k$-th and $k+1$-th excited states.
\end{itemize}

Please, note that this workload \ra{это KeyGenerator, а не workload} is infinite~--- the excited states are chosen in a cyclic manner, i.e., after the latest \emph{excited} state we get into the first one again through the dormant state. 

The TemporarySkewedKeyGenerator uses the following parameters:
\begin{itemize}
    \item $state$-$count$~--- a total number of excited states in the workload;
    \item $ht$~--- the default duration of an excited state (the duration is specified in number of iterations);
    \item $rt$~--- the default duration of a dormant state;
    \item during the $i$-th excited state, $p_i\%$ of operations are performed on $s_i\%$ of keys, and $100 - p_i\%$ of operations are performed on the rest of the set;
    \item $ht_i$~--- the duration of the $i$-th excited state, if necessary to set explicitly;
    \item $rt_i$~--- the duration of the dormant state after $i$-th excited state, if necessary to set explicitly.

\end{itemize}

\vspace{-0.2cm}
\subsection{Creakers and Wave KeyGenerator} \label{Creakers_and_Wave_KeyGenerator}
The Creakers and Wave Key Generator is based on the observation that the recently inserted keys are requested more often. They become obsolete over time and, consequently, are rarely requested. This workload \ra{это KeyGenerator, а не workload} is similar to workload D from YCSB with the Latest distribution: where more frequently accessed keys have higher probability to be an argument in the future.

For example, musical hits are listened to more often when they have just been released. Then, something new is released and the old is listened to less. To simulate this behaviour, the Creakers and Wave KeyGenerator has an entity Wave. The Wave is a subset of all keys from the range with a head and a tail. When it generates a new key, it adheres to the following rules:
\begin{itemize}
    \item \texttt{nextRemove()}~--- the Wave returns the key of the current tail and moves this tail by one;
    \item \texttt{nextInsert()}~--- we choose the key, next to the head of the Wave, and make it a new head;
    \item \texttt{nextGet()}~--- we select a key from the Wave with some specified distribution.
\end{itemize}

As a default, Wave uses the Zipfian Distribution, where the closer the key is to the head, the greater the probability of being selected.
Since there is a limit on the range of elements, to make the workload \ra{это KeyGenerator, а не workload} infinite the Wave moves in a cyclic manner over a working set of keys.
Besides the Wave there is an entity called Creakers. This is a subset of keys that are requested rarely but permanently. This entity is needed to check how well the data structure copes with such keys in the presence of rapidly growing and equally rapidly discarded keys from the Wave.


The Creakers and Wave KeyGenerator has the following parameters:
\begin{itemize}
    \item $cp\%$ of operations are performed on $cs\%$ of keys (Creakers entity), and $100 - cp\%$ are performed on the Wave entity;
    \item The Wave is initialized with $ws\%$ of keys; 
    \item $c$-$age$ get operations are performed on the Creakers entity during warmup before the benchmarking;
    \item $c$-$distribution$~--- a distribution of keys in the Creakers entity, the Uniform distribution by default;
    \item $w$-$distribution$~--- a mutable distribution of keys in the Wave entity: the Zipfian distribution with $\alpha=1$ by default.

\end{itemize}
All threads share the Wave head and tail.

One could argue that this workload \ra{это KeyGenerator, а не workload} is a copy of workload D from YCSB. There are two observations that highlight the differences. At first, our workload is infinite, while YCSB is predefined. There is no doubt that one can make YCSB work infinity, but the corresponding generator cannot generate a key with a time complexity of $O(1)$ in the runtime. This definitely affects the workload. Secondly, we added Creakers to our workload. Logically, nothing should really change~--- we just added some small permanent loads that always happen to the systems. However, it is a huge change for self-adjusting data structures~\cite{aksenov2023splay, afek2014cb}. The known implementations store a counter for each element, i.e., how many it was requested. The bigger the counter the faster the element is accessed. The problem occurs when elements are removed and lose their their whole counter. Thus, if we consider infinite execution, Creakers will have the largest counters, while the elements from the Wave will have smaller counters due to the insertion and removals. Obviously, this has a huge impact on the performance of the aforementioned index implementations.

\subsection{Leafs Handshake KeyGenerator} \label{Leafs_Handshake_KeyGenerator}
In the Leafs Handshake Key Generator, the key selection for an insert operation is based on the argument of the previous remove operation. Intuitively, the closer the key to the last removed one, the more probability that this key will be selected to insert. It accepts the following parameters:
\begin{itemize}
    \item $get$-$distribution$~--- a distribution of keys for a get operation (by default, the Uniform distribution);
    \item $remove$-$distribution$~--- a distribution of keys for a remove operation (by default, the Uniform distribution);    
    \item $insert$-$distribution$~--- a mutable distribution of keys for an insert operation (by default, the Zipfian distribution with $\alpha=1$).

\end{itemize}

\section{Implemented ThreadLoops}
\label{sec:threadloops}

\subsection{Default ThreadLoop}
The Default ThreadLoop selects the next operation with some fixed constant probability. It accepts the following parameters:
\begin{itemize}
    \item $ui\%$ of operations are insert operations;
    \item $ue\%$ of operations are remove operations.
    
\end{itemize}

\subsection{Temporary Operations ThreadLoop} \label{Temporary_Operations_ThreadLoop}
The Temporary Operations ThreadLoop selects the next operation depending on a time interval. It accepts the following parameters:
\begin{itemize}
    \item $temp$-$oper$-$count$~--- a number of different intervals;
    \item $ot_i$~--- a duration of $i$-th interval in the number of operations;
    \item $ui_i\%$ of operations are insert operations during the $i$-th interval;
    \item $ue_i\%$ of operations are remove operations during the $i$-th interval;

\end{itemize}
Time intervals are repeated cyclically, giving an infinite workload \ra{Это ThreadLoop, а не Workload}.

\section{Separation of Binary Search Trees}
\label{sec:separator}

Our benchmark is written in Java and it is available at the link~\url{https://github.com/Mr-Ravil/synchrobench}.

We consider three commonly-known binary search tree (BST) implementations in Java. Our main goal is to show that using our workloads we can show any possible different relative performance, i.e., there is no obvious winner. There are six possibilities, since we have only three data structures.

\vspace{-0.2cm}
\subsection{Implementations}
We take the three most effective binary search tree (BST) implementations written in Java:
\begin{itemize}
    \item A BCCO BST by Bronson et al.~\cite{bronson2010practical}.
    \item A Contention-Friendly (CF) BST by Gramoli et al.~\cite{crain2013contention}.
    \item A Concurrency-Optimal (CO) BST by Aksenov et al.~\cite{aksenov2017concurrency}. 
\end{itemize}

All of them are partially-external and the the main difference between them is how they handle physical removal and rotation.
In the BCCO-BST
a working thread always removes nodes physically and rotates subtrees, if necessary.
In the CF-BST
a working thread makes only logical removals. However, there is a special daemon thread responsible for physical removals and rotations.
In the CO-BST
a thread performs physical removal immediately but does not perform rotations at all.

Looking on the experiments from the previous papers using the workloads from Synchrobench, e.g., \cite{aksenov2017concurrency}, one could decide that Concurrency-Optimal is the superior one, the runner-up is Contention-Friendly, and the last one is BCCO. Obviously, it should not be the case on all the possible workloads because: 1) CO does not make rotations at all, leading to longer traversals; 2) CF should perform worse on the larger trees, since the daemon thread could not catch up with changes.
We show that it is indeed the case: depending on the workload we can see different relative performance. Prior to that, we must introduce a small change to the Concurrency-Friendly implementation that sometimes improves performance by a large amount.

\subsubsection{Concurrency-Friendly Improvement}

During the experiments, we found that it is possible to improve the daemon thread algorithm.
In the Synchrobench implementation, the thread attempts to remove the node first, and, then, runs recursively.
However, we can go recursively to children, and only after that we can restructure the node itself.
    
Thus, if there are $N$ elements in the tree and all of them are logically removed, in our algorithm the daemon requires one call of a recursive restructure function. While the daemon thread from the Synchrobench version requires $\log N$ calls leading to $\sum\limits_{i=1}^{\log N} (2^i - 1) = 2N - \log N - 2$ operations. We could conclude that the daemon thread algorithm from Synchrobench is almost twice as slow, which may affect the overall throughput. In the experiments, we call our implementation as Fixed Concurrency-Friendly version.

\subsection{Experimental Results}
We evaluateed workloads on two-processor Intel Xeon Gold 6240R with $24$ cores each, giving $48$ cores in total. 
We show results starting from 16 working threads, since this is the most interesting part. Each plot has four graphs: blue (CO), yellow \ra{это цвет оранге, а не желтый O\_о } (Fixed CF), green (CF), and red (BCCO). OX axis represents the number of used threads, and OY axis represents the throughput, i.e., the number of operations performed in one second.
Each point in plots is the result of an experiment performed $10$ times for $10$ seconds.

\subsubsection{Uniform and Zipfian Workloads}
Uniform and Zipfian Workloads are the most common workloads. We run it with the following parameters: range is $10^4$ (for Uniform) and $10^5$ (for Zipfian) and update ratio is $20\%$ (for Uniform) and $5\%$ (for Zipfian). $insert$ ratio is equal to $remove$ ratio.

\begin{figure}
    \centering
    \includegraphics[width=\linewidth]{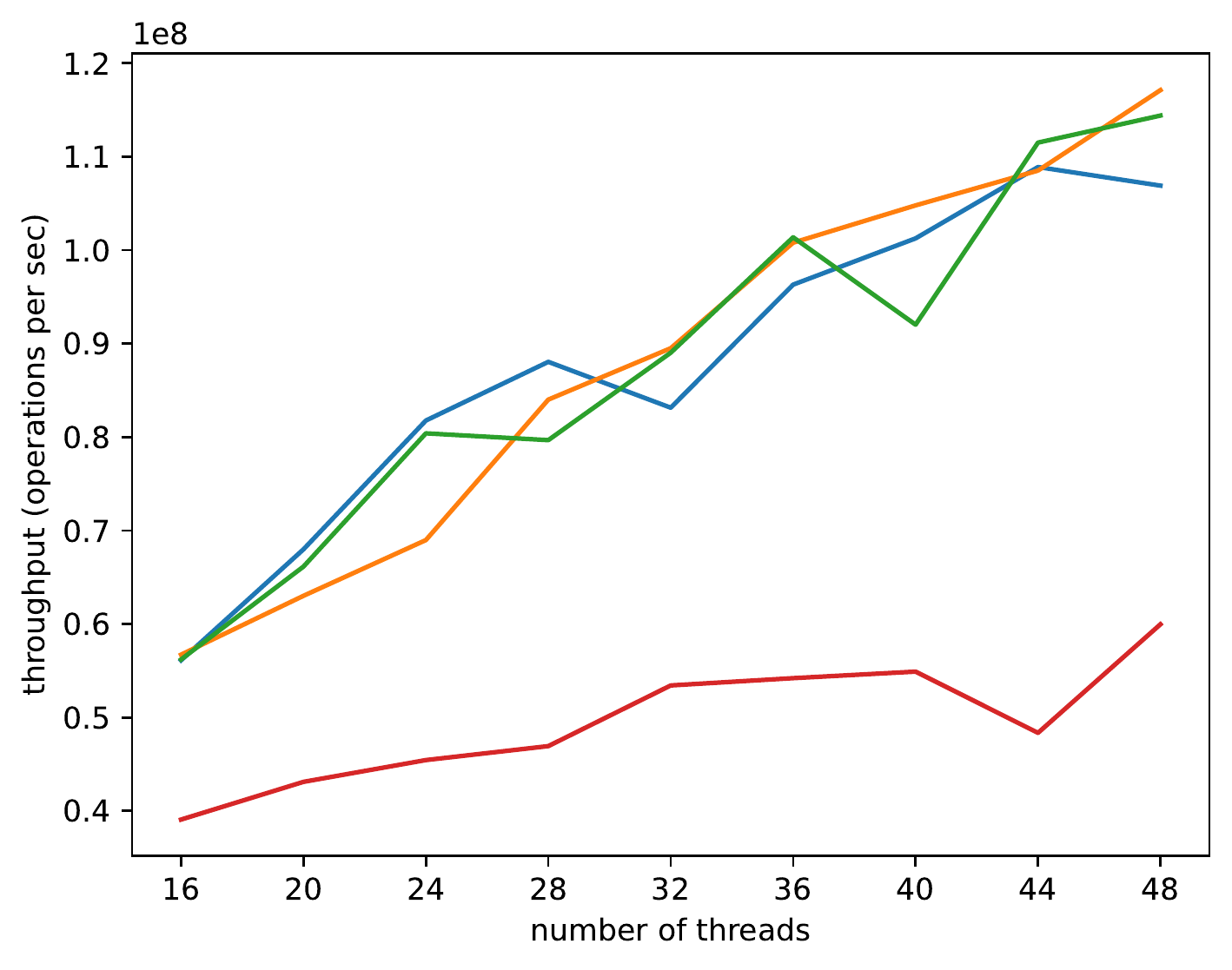}   
    \caption{Uniform Workload with range $10^4$ and update ratio $20\%$}
    \label{fig:Uniform_Experiment}
\end{figure}

\begin{figure}
    \centering
    \includegraphics[width=\linewidth]{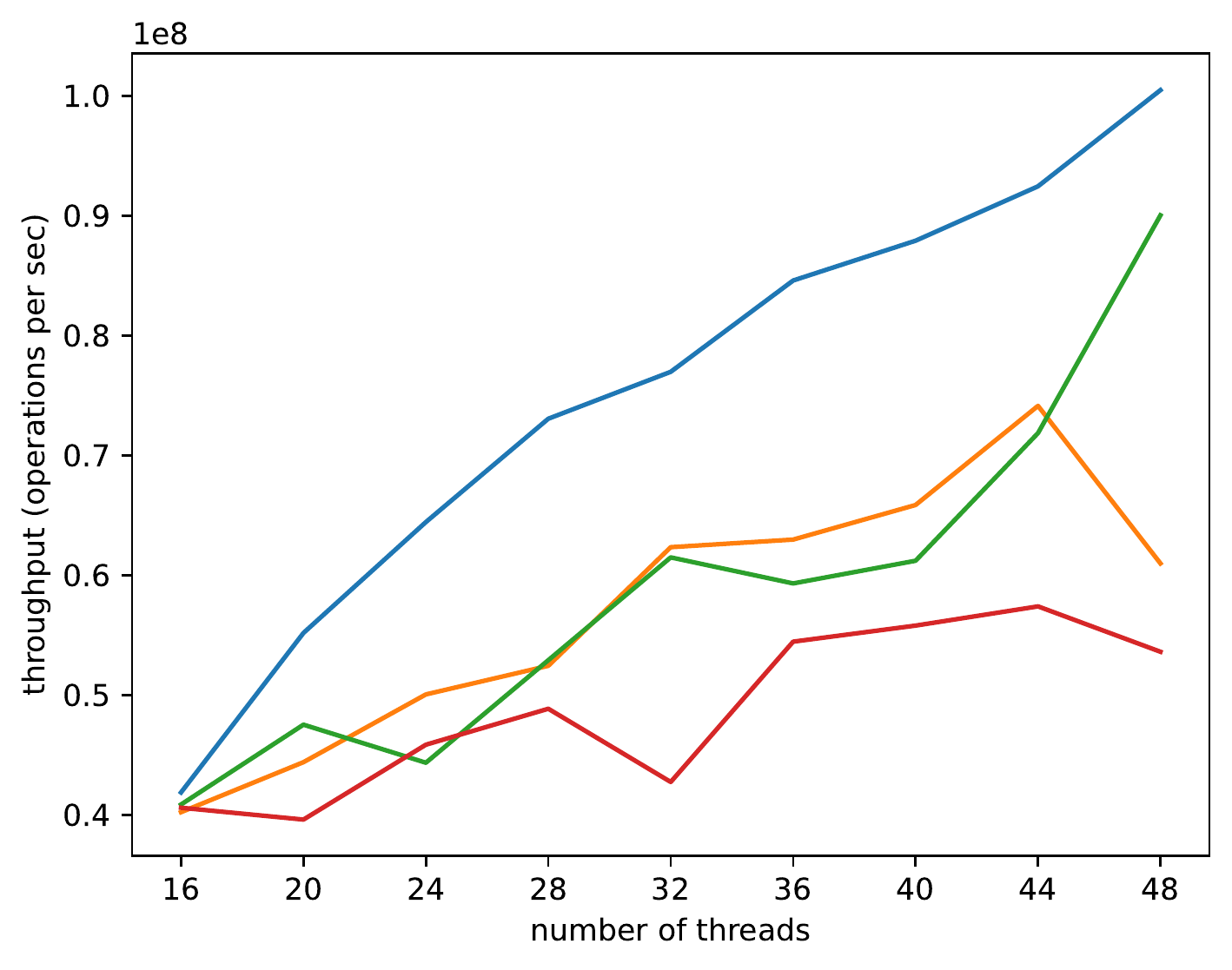}
    \caption{Zipfian Workload with $\alpha=1$, range $10^5$ and update ratio $5\%$.}
    \label{fig:Zipfian_Experiment}
\end{figure}

The results can be seen on Figures~\ref{fig:Uniform_Experiment}~and~\ref{fig:Zipfian_Experiment}.
Both workloads have the same resulting pattern: CO-CF-B, i.e., the first~-- Concurrency-Optimal, the second~--- Concurrency-Friendly, and the third~--- BCCO.



\subsubsection{Infinite Leafs Handshake Workload}
Since BCCO-BST always rotates and removes physically, one expects it to be too far behind in terms of throughput. But that does not mean that it works badly on any workload. We present the Infinite Leafs Handshake Workload that is based on the Leafs Handshake KeyGenerator~\ref{Leafs_Handshake_KeyGenerator} and the Temporary Operations ThreadLoop~\ref{Temporary_Operations_ThreadLoop}. It has three time intervals: 1) the filling interval~--- there are more insertions than removal; 2) the read interval~--- there are only read operations; 3) the cleaning interval~--- there are more removals than insertions.

During the filling interval, for one removed node, two new neighbors are inserted.
During the reading interval, the working threads just read the keys uniformly at random.
During the cleaning interval, the working threads remove nodes uniformly at random from the tree to restore its original size. This phase is added to make the workload infinite.

\begin{figure}
    \centering
    \includegraphics[width=\linewidth]{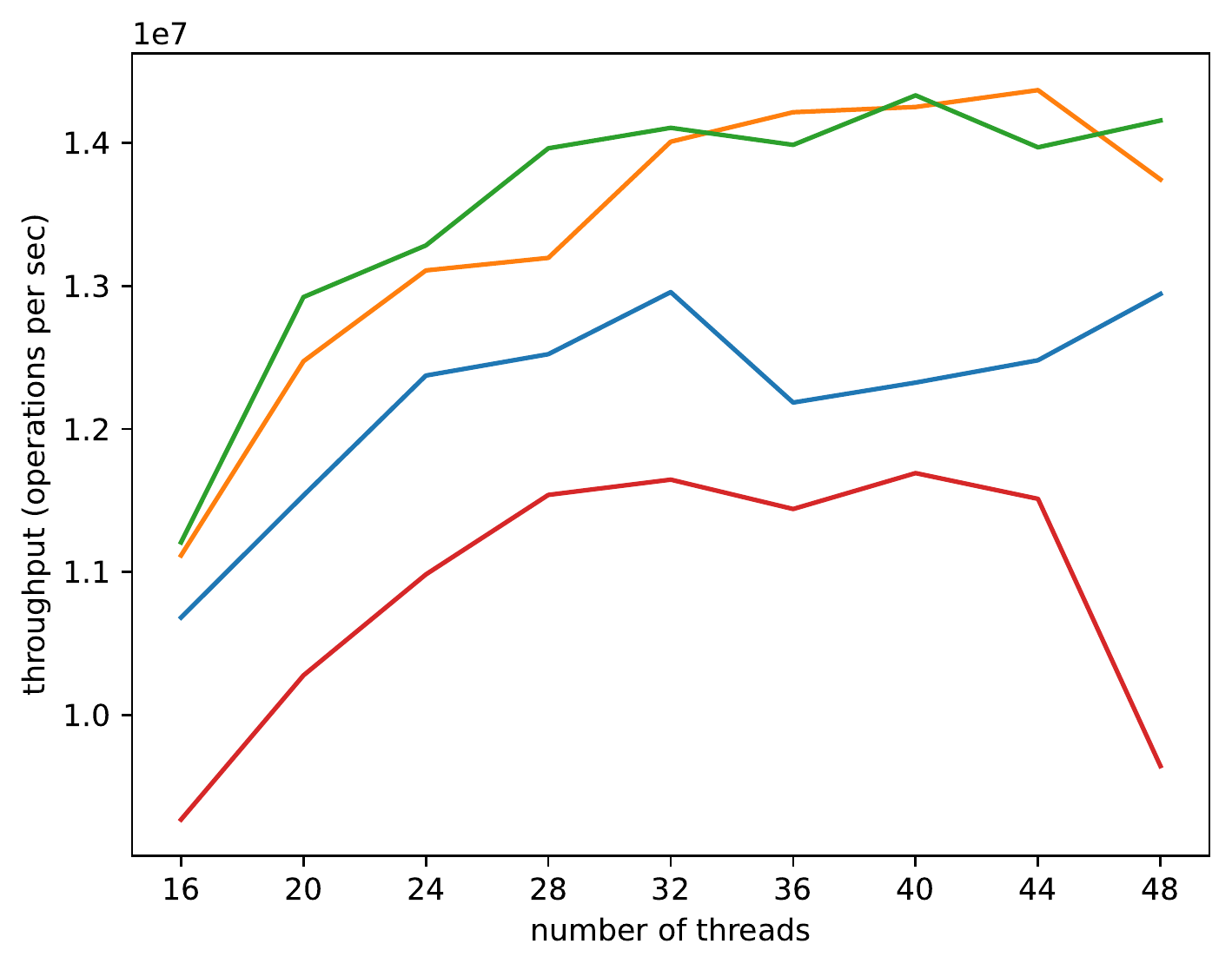}
    \caption{Infinite Leafs Handshake Workload with range $10^5$.}
    \label{fig:Little_Infinite_Leafs_Handshake_Experiment}
\end{figure}

The first experiment (Figure~\ref{fig:Little_Infinite_Leafs_Handshake_Experiment})  is run with parameters: range is $10^5$, temp-oper-count is $3$; $ot_0 = ot_2 = 10000$, $ot_1=5000$; $ui_0 = ue_2 = 60\%$; $ui_2 = ue_0 = 40\%$; $ui_1 = ue_1 = 0\%$; get and remove distributions are uniform distributions; insert distribution is a Zipfian distribution with $\alpha=2$. It gives us the relative performance of CF-CO-B.

\begin{figure}
    \centering
    \includegraphics[width=\linewidth]{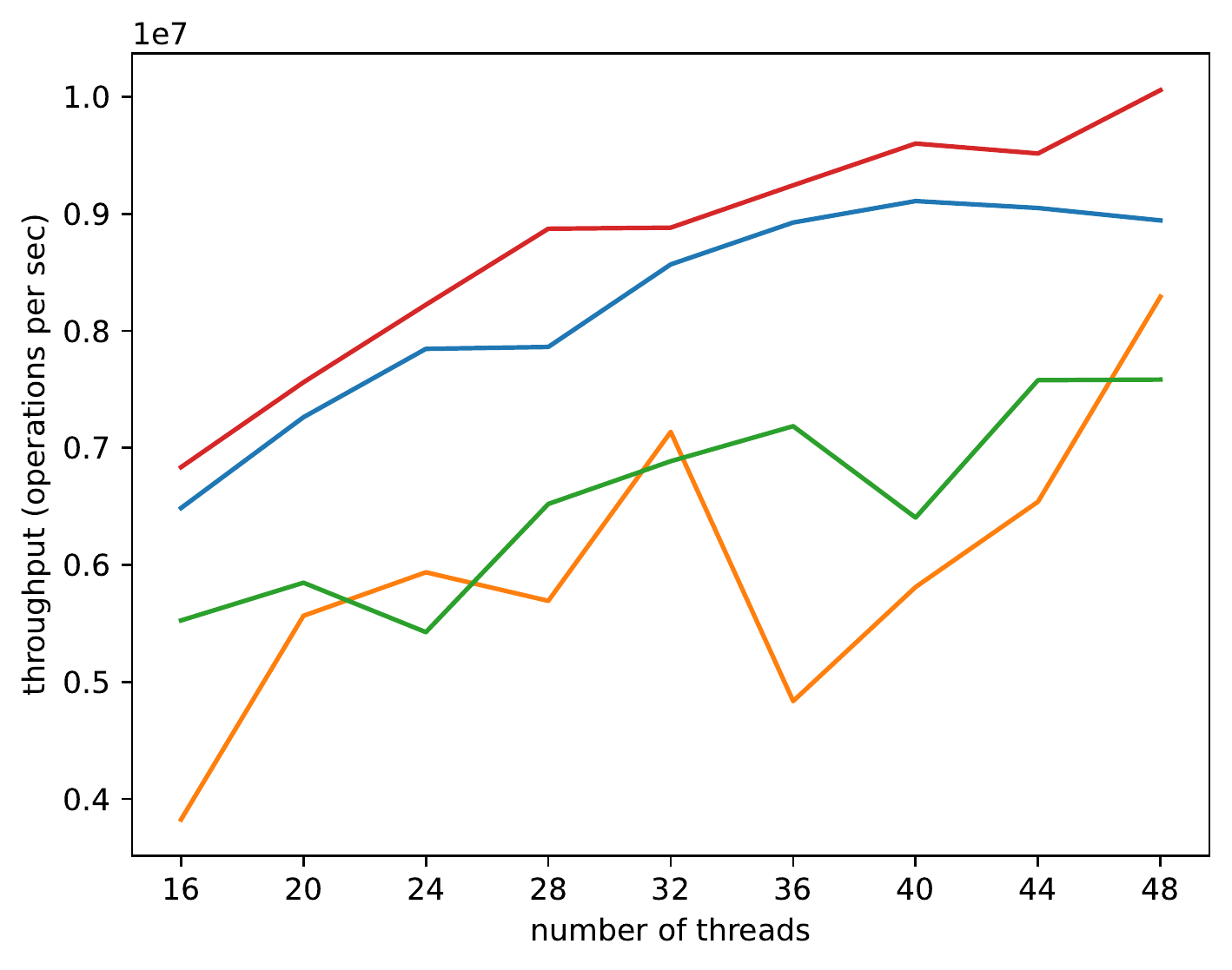}
    \caption{Infinite Leafs Handshake Workload with range $10^7$.}
    \label{fig:Infinite_Leafs_Handshake_Experiment}
\end{figure}

The second experiment (Figure~\ref{fig:Infinite_Leafs_Handshake_Experiment}) is run with parameters: range is $10^7$; temp-oper-count is $2$; $ot_0 = ot_1 = 20000$; $ui_0 = ue_1 = 90\%$; $ui_1 = ue_0 = 10\%$; get and remove distributions are Uniform; and insert-distribution is a Zipfian distribution with $\alpha=0.99$. It gives us the relative of performance B-CO-CF.

\begin{figure}
    \centering
    \includegraphics[width=\linewidth]{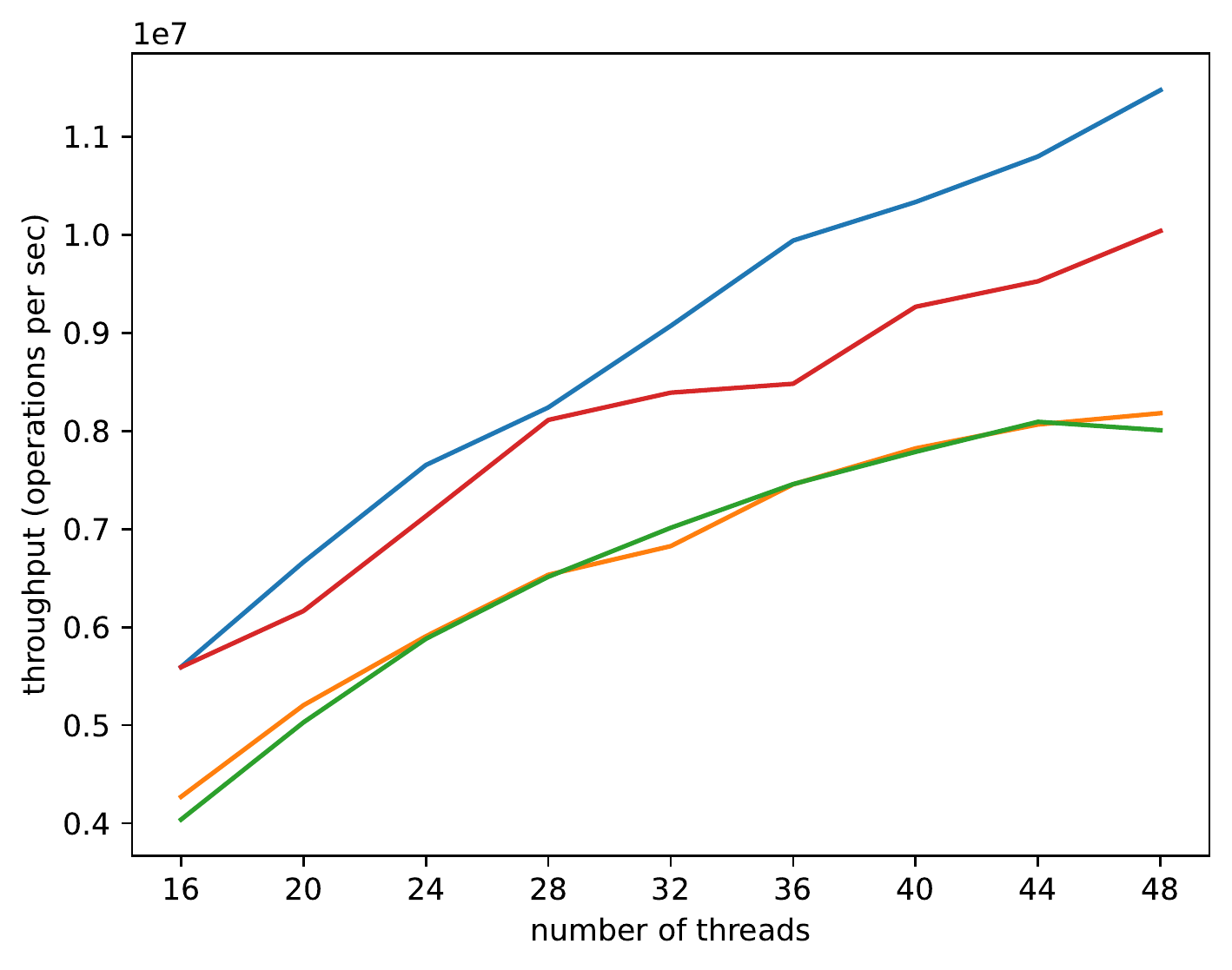}
    \caption{Infinite Leafs Handshake Workload with range $10^8$}
    \label{fig:Big_Infinite_Leafs_Handshake_Experiment}
\end{figure}

The third experiment (Figure~\ref{fig:Big_Infinite_Leafs_Handshake_Experiment}) is run with parameters: range is $10^8$; temp-oper-count is $3$; $ot_0 = ot_1 = ot_2 = 100000$; $ui_0 = ue_2 = 80\%$; $ui_2 = ue_0 = 20\%$; $ui_1 = ue_1 = 0\%$; get and remove distributions are Uniform; insert distribution is a Zipfian distribution with $\alpha=0.99$. It gives us the relative performance of CO-B-CF.

From the results, it can be seen that CF-BST lags far behind at large ranges (Figures~\ref{fig:Infinite_Leafs_Handshake_Experiment} and~\ref{fig:Big_Infinite_Leafs_Handshake_Experiment}), while at small ranges it behaves about the same or better than others (Figure~\ref{fig:Little_Infinite_Leafs_Handshake_Experiment}).This happens because the daemon thread does not catch up with remove operations and the tree has a longer traversal length.
Also, CO-BST and BCCO-BST behave differently at large ranges. The less skew between \texttt{insert} and \texttt{remove} operations, the better CO-BST handles the workload. This is because with more skew there are more chances that the tree will grow non-uniformly, which makes the CO-BST perform poorly.

\subsubsection{Non-shuffle Wave Workload}
The Non-shuffle Wave Workload is based on Creakers and Wave KeyGenerator~\ref{Creakers_and_Wave_KeyGenerator} without the Creakers and the shuffle in KeyGeneratorData. 
This workload adds new keys to the edge of the tree disrupting the balance. That leads to the performance problems of poorly balanced trees.

\begin{figure}
    \centering
    \includegraphics[width=\linewidth]{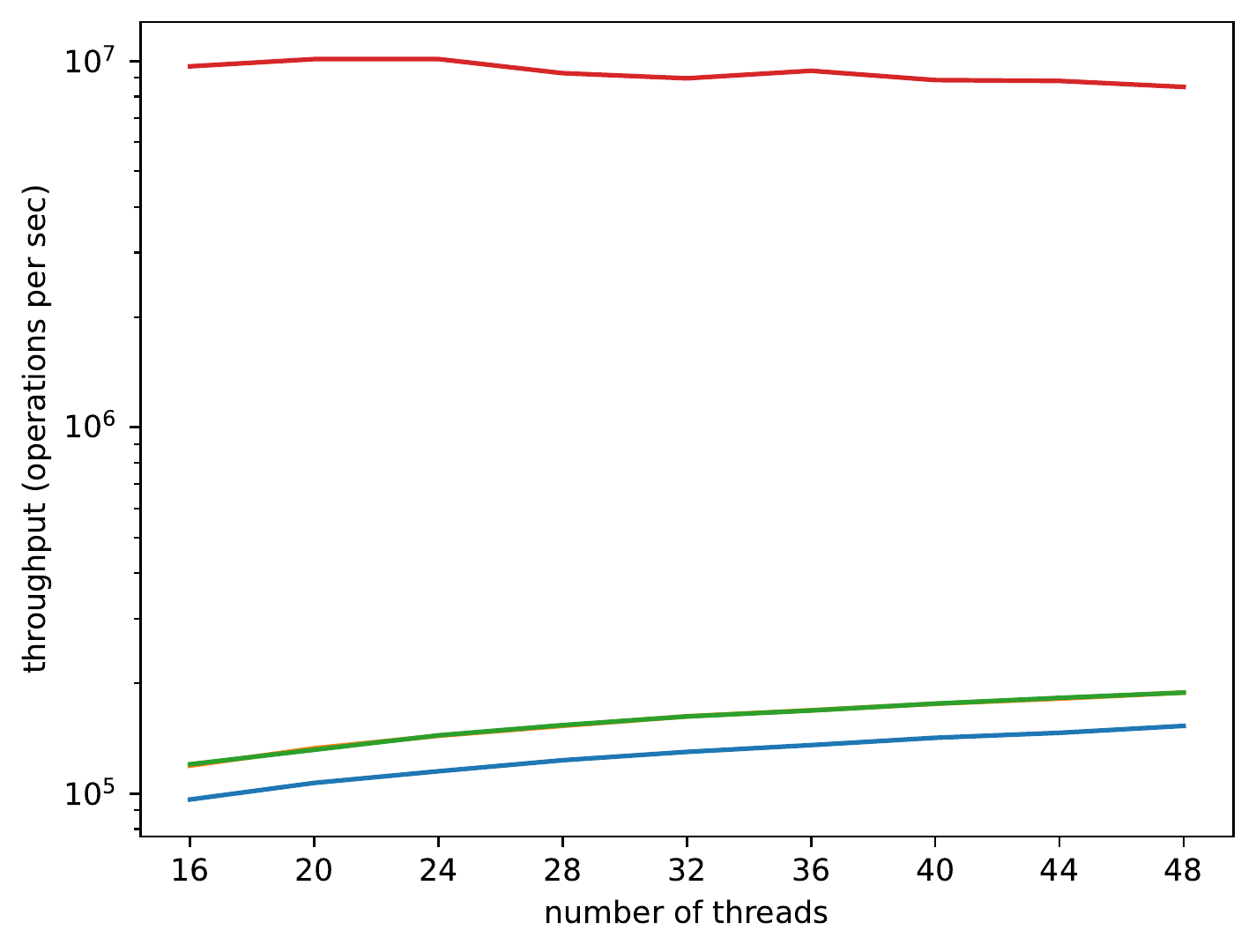}
    \caption{Non-shuffle Wave Workload with range $10^6$.}
    \label{fig:Non_shuffle_Wave_Experiment}
\end{figure}

The first experiment (Figure~\ref{fig:Non_shuffle_Wave_Experiment}) is run with parameters: range is $10^6$; $ws = 20\%$; update ratio is $5\%$; $w$-$distribution$~--- Zipfian distribution with $\alpha=1$; and $cp = 0\%$. It gives us the relative performance of CF-B-CO.

This result is again related to the fact that the BCCO-BST always rotates which is not the case of CF-BST and CO-BST.

\begin{figure}
    \centering
    \includegraphics[width=\linewidth]{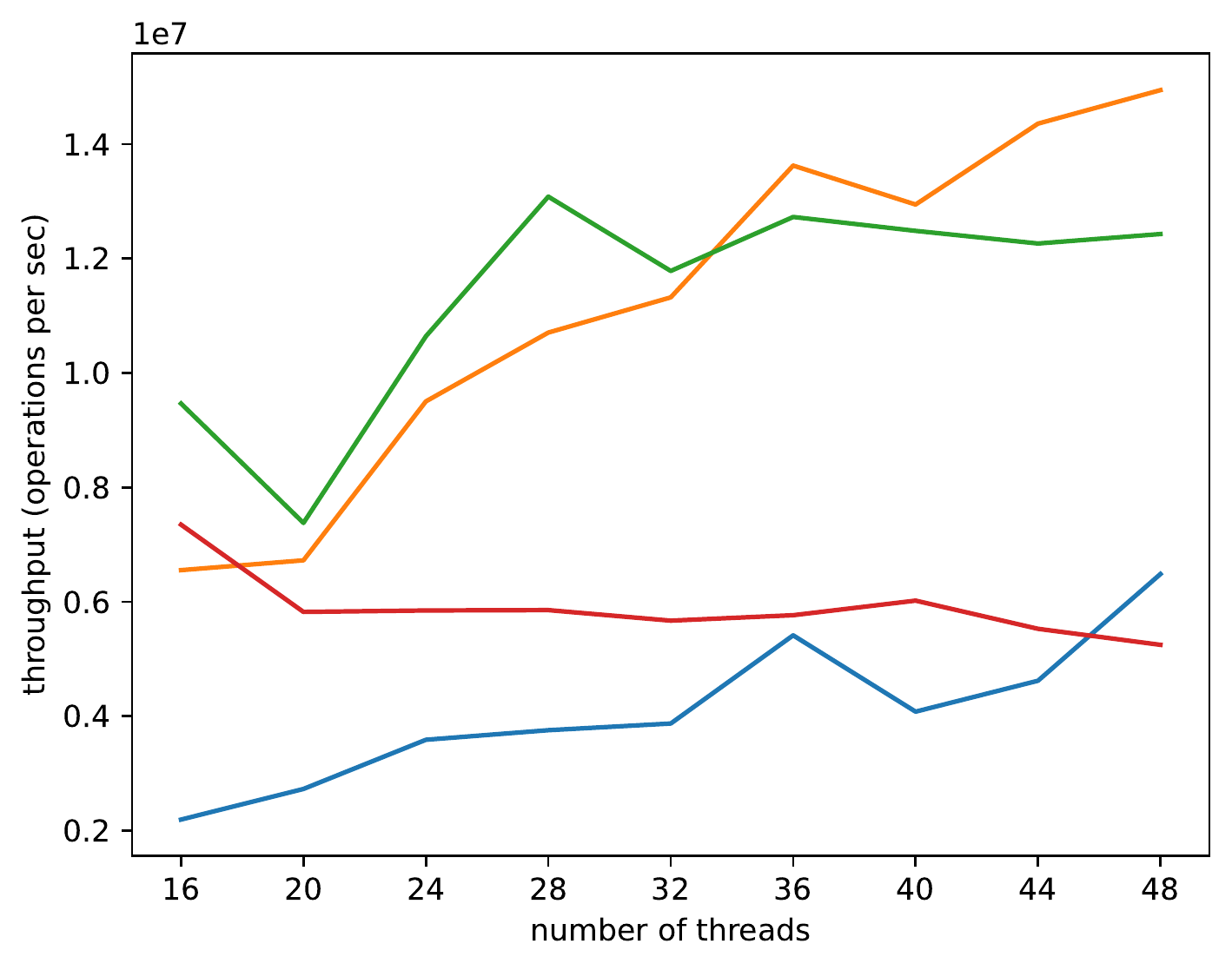}
    \caption{Non-shuffle Wave Workload with range $5\cdot10^3$.}
    \label{fig:Little_Non_shuffle_Wave_Experiment}
\end{figure}

The second experiment (Figure~\ref{fig:Little_Non_shuffle_Wave_Experiment}) is run with parameters: range is $5\cdot10^3$; $ws = 10\%$; write ratio is $20\%$; $w$-$distribution$~--- Zipfian distribution with $\alpha=1$; $cp = 0\%$. It gives us the relative performance of B-CF-CO.

This result is related to the fact that the tree size is quite small, which is why the daemon thread in CF-BST manages to balance the tree quickly.


\subsection{Summary}
As a result we obtained all six relative performances of three data structures: B-CF-CO on Figure~\ref{fig:Little_Non_shuffle_Wave_Experiment}, B-CO-CF on Figure~\ref{fig:Infinite_Leafs_Handshake_Experiment}, CF-B-CO on Figure~\ref{fig:Non_shuffle_Wave_Experiment}, CF-CO-B on Figure~\ref{fig:Little_Infinite_Leafs_Handshake_Experiment}, CO-B-CF on Figure~\ref{fig:Big_Infinite_Leafs_Handshake_Experiment}, and CO-CF-B on Figures~\ref{fig:Uniform_Experiment}~and~\ref{fig:Zipfian_Experiment}.

\section{Conclusion}
In this work, we presented a new benchmarking suite that tries to solve issues that appeared in other known suites. There we introduced several workloads inspired by real-life. At the end, we showed the most important observation of the paper. The benchmarking process of different data structures is a really complicated problem and it should be done with care: depending on the workload one can get totally ``contradictory'' results that can affect the conclusion about the performance.

As for future work, we want to polish the suite for Java and prepare the similar suite for C++.

\pagebreak

\bibliographystyle{ACM-Reference-Format}
\bibliography{references}



\end{document}